\documentclass[12pt,preprint]{aastex}

\def\lessim{\mathrel{\hbox{\rlap{\hbox{\lower4pt\hbox{$\sim$}}}\hbox{$<$}}}}
\def\grtsim{\mathrel{\hbox{\rlap{\hbox{\lower4pt\hbox{$\sim$}}}\hbox{$>$}}}}

\shorttitle{Nova M31N~2007-11d}
\shortauthors{Shafter et al.}

\begin{document}


\title{M31N~2007-11d: A Slowly-Rising, Luminous Nova in M31}


\author{A. W. Shafter\altaffilmark{1}, A. Rau\altaffilmark{2}, R. M. Quimby\altaffilmark{2}, M. M. Kasliwal\altaffilmark{2,3}, M. F. Bode\altaffilmark{4}, M. J. Darnley\altaffilmark{4}, and K. A. Misselt\altaffilmark{5}}
\altaffiltext{1}{Department of Astronomy, San Diego State University,
    San Diego, CA 92182}
\altaffiltext{2}{Division of Physics, Mathematics, and Astronomy, 105-24, California Institute of Technology, Pasadena, CA 91125}
\altaffiltext{3}{Hale Fellow, Gordon and Betty Moore Foundation}
\altaffiltext{4}{Astrophysics Research Institute, Liverpool John Moores University, Birkenhead CH41 1LD, UK}
\altaffiltext{5}{Steward Observatory, University of Arizona, Tucson, AZ 85721}



\begin{abstract}
We report a series of extensive photometric and
spectroscopic observations of the luminous M31 nova M31N~2007-11d.
Our photometric observations coupled with previous measurements
show that the nova took at least four days
to reach peak brightness at $R\simeq14.9$ on 20~Nov~2007~UT. After
reaching maximum, the time for the nova to
decline 2 and 3 magnitudes from maximum light ($t_2$ and $t_3$)
was $\sim9.5$ and $\sim13$~days, respectively,
establishing that M31N~2007-11d was a moderately fast declining nova.
During the nova's evolution a total of three
spectra were obtained. The first spectrum was obtained one day
after maximum light (5 days post-discovery), followed
by two additional spectra taken on the decline at
two and three weeks post-maximum.
The initial spectrum reveals narrow
Balmer and Fe~II emission with P~Cygni profiles superimposed
on a blue continuum.
These data along with the spectra obtained on the subsequent decline 
clearly establish that M31N~2007-11d belongs to
the Fe~II spectroscopic class. The properties of M31N~2007-11d
are discussed within the context of other luminous
novae in M31, the Galaxy, and the LMC.
Overall, M31N~2007-11d appears to be remarkably similar
to Nova LMC~1991, which was
another bright, slowly-rising, Fe~II nova.
A comparison of the
available data for luminous extragalactic novae suggest that the
$\grtsim4$ day rise to maximum light seen in M31N~2007-11d
may not be unusual, and that the rise times of
luminous Galactic novae, usually assumed to be $\lessim2$ days,
may have been underestimated.

\end{abstract}




\section{Introduction}

Classical Novae are a sub-class of cataclysmic variable
systems where a Roche-lobe-filling
star (typically a cool, near-main-sequence star) transfers
mass to a white dwarf companion (Warner 1995, 2008). Eventually,
a thermonuclear runaway (TNR) in the accreted material
ensues, which drives substantial
mass loss from the system, and leads to the nova eruption
(e.g. Starrfield et al. 2008).
Absolute magnitudes as bright as M$_V \simeq -10$ have been observed
at the peak of eruption.
Their high luminosities and high rates of occurrence
($\sim30$~yr$^{-1}$ in a galaxy like the Milky Way [Shafter 2002]),
make novae powerful
probes of the evolution of binary systems in different (extragalactic) stellar
populations.
The most throughly studied extragalactic system is M31, where
more than 700 novae have been discovered since
Hubble (1929) began his pioneering work
in the early 20th century (e.g. see Darnley et al. 2006; Pietsch
et al. 2007; Shafter 2008, and references therein).

In November of 2007, Nakano (2007) reported the discovery
of a slowly-rising, and
particularly luminous nova in M31. The nova, M31~2007-11d,
was initially detected at $m=17.7$ (unfiltered) on 2007~Nov~16.51
before reaching $m=14.9$ on Nov~20.385. It was discovered
independently by Quimby et al. (2007) as part of the ROTSE~IIIb patrol,
and a finding chart based on these data showing the position
of M31~2007-11d within M31 is shown in Figure~1.
According to the compilation of M31 novae by Pietsch et al. (2007),
only six (including M31~2007-11d)
of the more than 700 novae recorded in M31 have
been observed with $m<15$.
Shortly after it became clear that M31~2007-11d was an unusually
luminous nova, we initiated a series of photometric and
spectroscopic observations to follow its subsequent evolution.
Here, we report the results of this campaign.

\section{Observations}

\subsection{Photometry}

We extensively monitored M31~2007-11d with the Palomar 60-inch
telescope (P60; Cenko et al. 2006), the
Faulkes Telescope North (FTN; Burgdorf et al.
2007), and the 2-m Liverpool Telescope (LT; Steele et al. 2004).
The P60 observations (see Table~1) started on
20.4 Nov 2007 UT, four days post-discovery and  continued for 21 days
until the source faded beyond detectability.  Data were reduced within
the  IRAF\footnote{IRAF  is   distributed  by  the  National  Optical
Astronomy  Observatory,  which  is  operated by  the  Association  for
Research  in Astronomy,  Inc.   under cooperative  agreement with  the
National Science Foundation.} environment. Imaging was obtained in the
$g, r,  i'$ and  $z'$ pass bands  and photometrically  calibrated with
respect to SDSS.
The FTN and LT data (see Table~2)
were taken through $B$, $V$ and Sloan $i'$ filters using
the HawkCam and RATCam instruments on the FTN and LT, respectively. These
data were reduced using standard routines within IRAF and Starlink, and
calibrated against standard stars from Landolt (1992).

In addition to our targeted observations,
we have been able to supplement our photometric coverage of M31~2007-11d
by extracting unfiltered magnitudes from
the on-going monitoring of M31 taken as part of the ROTSE-IIIb program (see
Table~3), and with photometric
measurements made on the night of 25~Nov~2007 using the Mount Laguna
Observatory (MLO) 1-m reflector. For the latter observations,
the field of the nova was imaged with a Loral CCD
detector through $B, V$, and $R$ filters. As with the
FTN and LT data, the MLO observations were reduced with
standard routines in IRAF, and calibrated with standard stars
from Landolt (1992). M31N~2007-11d had
faded to $V=16.31$ by the time of the MLO observations, and
was characterized at that time by $B-V=0.22$ and $V-R=0.27$.

Our comprehensive light curve for M31~2007-11d is shown in Figure~2.
In addition to the extensive photometry on the decline from maximum,
we have also
included some early unfiltered CCD measurements by
K. Nishiyama and F. Kabashima as communicated by Nakano\footnote{
http://www.cfa.harvard.edu/iau/CBAT\_M31.html},
which are useful in establishing the time of maximum light,
and in constraining the rise time of the eruption. The time
interval between the initial discovery
on 2007~Nov~16.51 and the time the nova reached
peak brightness on Nov~20.4 allows us to
set a lower limit of $\sim4$ days on the rise to maximum. On the subsequent
decline, the light curve indicates that M31~2007-11d was a relatively
``fast" nova, with times to decline by two and three magnitudes
($V$-band) from maximum light ($t_2[V]$ and $t_3[V]$)
of 9.5 and 13~days, respectively. The corresponding decline rate
of $\sim0.22\pm0.01$~mag~d$^{-1}$ is consistent with those of other
luminous novae in M31 (Capaccioli et al. 1989).

\subsection{Spectroscopy}

Our spectroscopic observations of M31~2007-11d were obtained
with both the Hobby-Eberly Telescope (HET) and the Keck-I Telescope.
The HET observations were made
with the Low Resolution Spectrograph (LRS; Hill et al. 1998)
as part of an on-going spectroscopic survey of novae in M31.
We used the $g1$ grating with a
1.0$''$ slit and the GG385 blocking filter, which covers
4150-11000\,\AA\ with a resolution of $R\sim 300$, although we limit
our analysis to the 4150-8900\,\AA\ range where the effects of order
overlap are minimal.
The Keck-I data were taken
in long-slit mode with
the Low Resolution  Imaging Spectrograph (LRIS; Oke 1995).
LRIS is a dual beam instrument
in which the light is split by a dichroic, and spectra can be obtained
in two channels simultaneously.  We obtained 600\,s exposures with the
400/3400  (dispersion of  $1.\!\!^{\prime\prime}$09 per  pixel) grating
and  the 400/8500  $(1.\!\!^{\prime\prime}$86 per  pixel) grism  on the
blue  and red  sides,  respectively.
Both the HET and Keck spectra were reduced using standard IRAF routines,
and are shown in Figures~3 and~4, respectively.
For reference with the light curve,
the dates when these spectra were obtained are
indicated in Figure~2.

Our initial HET spectrum was taken on 21~Nov~2007~UT,
roughly five days after the nova began to brighten ($\sim$1~day post
maximum) when M31N~2007-11d was at $V\sim15.5$.
The spectrum is dominated by Balmer, Fe~II, and
P~Cygni emission features (FWHM $\simeq 1600$ km$^{-1}$)
superimposed on a very blue continuum. These features
are characteristic of a
nova caught at maximum light, but are unusual for a
nova five days post discovery. This discrepancy can
be attributed to the relatively slow rise to maximum
exhibited by M31N~2007-11d.
Overall, the spectrum is consistent with
a classification as an Fe~II nova in
the system of Williams (1992).

A follow-up HET spectrum, also shown in Figure~3,
was obtained on 04~Dec~2007, 18 days post eruption ($\sim14$ days after
maximum).
Our photometry indicates that the nova had declined
to $V\sim19$, more than 4 magnitudes from maximum light,
and was fading rapidly.
As is typical of the early spectral
development of novae, the spectrum changed dramatically
from that near maximum light,
with the fading of the strong blue continuum and the
disappearance of the P~Cygni profiles. The spectrum became
dominated by strong Balmer (FWHM $\sim$ 2300~km~s$^{-1}$),
Fe~II and O~I lines in emission, confirming the identification
of M31N~2007-11d as an Fe~II-type nova.

Our final spectrum (Figure~4) was obtained
on 11.5 Dec 2007 UT after M31N~2007-11d
had faded to well below $V=21$. By this time
most emission features had faded significantly, with
only residual H$\alpha$ emission (FWHM $\sim$ 1800~km~s$^{-1}$)
being clearly detected. 

\section{Discussion}

\subsection{The Observed M31 Nova Luminosity Distribution}

Through the end of 2007, a total of 779 novae have been discovered
in M31 since Hubble's observations began nearly a century ago
(Hubble 1929).
Of these, 732 novae have measured magnitudes (Pietsch et al. 2007).
The interpretation of these data is complicated by the
fact that not all novae were observed on the same photometric
system, or in the same bandpass. Most of the early observations yielded
photographic magnitudes, with much of the later data yielding either
unfiltered or H$\alpha$ magnitudes. Typically,
observations have been reported in one or more of the
$m_{pg}, U, B, V, R, I, W$ (unfiltered), and H$\alpha$ bandpasses.

In order to better compare the data, we have attempted to
convert all individual photometric
observations to a reddening-free visual magnitude, $V_0$.
We began this process by correcting all observations for foreground
extinction using a reddening of $E(B-V)=0.062$ along the line of sight to M31
found by Schlegel et al. (1998).\footnote{We have made no
attempt to correct for extinction internal to M31, which is highly
variable and difficult to quantify based on the data available for
individual novae.}
From this, we estimate $A_U=0.30$, $A_B=A_{pg}=0.25$,
$A_V=0.20$, $A_R=A_{H\alpha}=A_W=0.15$, and $A_I=0.10$, where the
H$\alpha$ and $W$ (unfiltered) CCD magnitudes have been treated as
$R$-band measurements since the effective wavelengths are similar.
After correcting for extinction, we converted
all observations to the visual band by adopting representative
colors for a typical nova shortly after eruption.
Van den Bergh \& Younger (1986) have shown that the mean colors
of Galactic novae are given by $<$$B-V$$>_0=0.23\pm0.06$
at maximum light, and $<$$B-V$$>_0=-0.02\pm0.04$ by the time the
average nova had faded two magnitudes below maximum. Since
we expect that the typical M31 nova will have faded somewhat by the time of
discovery, here we adopt $<$$B-V$$>_0\simeq0.15$ as a plausible estimate
for the color at the time of the photometric measurements. This value
is consistent with
the de-reddened color of M31N~2007-11d, $(B-V)_0=0.16$, based on our MLO
observations obtained $\sim5$ days post maximum.
According to the data compiled by Johnson (1966), this is the
approximate color of an A5V star, which is also characterized by
$(U-B)_0=0.11$, $(V-R)_0=0.16$, and $(R-I)_0=0.06$.
Since the energy distribution of a nova near maximum light is similar
to a blackbody, we can use these color indicies to convert all
M31 nova observations to a de-reddened $V$ magnitude
(photographic magnitudes were initially converted
to $B_0$ magnitudes using the relationship,
$[B-m_{pg}]_0=0.27-0.19~[B-V]_0$ from Arp [1961]).

Given the distance of M31 ($\mu_0=24.38$; Freedman et al. 2001),
our values of $V_0$ result in
the absolute magnitude distribution shown in
Figure~5, with
the position of M31~2007-11d ($M_V\simeq-9.5$) indicated
by the arrow. It is important to note that
since the photometric measurements of M31 novae were
not all obtained precisely at maximum light,
this distribution is only an approximation to the true maximum
magnitude distribution of novae in M31.
Nevertheless, it is likely that the majority of
systems were observed near maximum, particularly for novae
at the brighter end of the distribution.
Clearly, M31~2007-11d, which we estimate
reached an absolute magnitude at maximum light of $M_V\simeq-9.5$,
is one of the
brightest novae to erupt in M31 over the past century.

\subsection{Evidence for Two Nova Populations?}

There has been a growing body of evidence from both
Galactic and extragalactic observations that there
may exist two separate populations of novae
(e.g., see Shafter 2008, and references therein).
This conjecture is based primarily on three principal findings:
(1) The frequency distribution
of nova magnitudes at maximum light from the
Arp (1956) M31 nova survey is bimodal, possibly as a result of two
distinct populations of progenitor systems
(There is also some evidence for bimodality in the Rosino
[1973] M31 survey, but it is less compelling),
(2) Galactic observations suggesting that novae
associated with the disk may be on average faster and more luminous
than novae thought to be associated with the bulge
(e.g. Duerbeck 1990; Della Valle et al. 1992), (3) The distribution of
the rates of decline for novae in the LMC suggesting that these
novae are on average ``faster" than those in the bulge of M31
(Della Valle \& Duerbeck 1993), and (4) The realization
that novae can be divided into two principal classes based upon
their spectroscopic properties (Williams 1992).
Novae with prominent Fe~II lines (the Fe~II novae) usually show
P~Cygni absorption profiles, and tend to evolve more slowly, have lower
expansion velocities, have a lower level of ionization,
compared with novae that
exhibit strong lines of He and N (the He/N novae). Later,
Della Valle \& Livio (1998) noted that Galactic novae that were
classified as He/N were concentrated near the Galactic plane, and tended to
be faster, and more luminous compared with their Fe~II counterparts.

Assuming that the bimodal magnitude distribution seen by Arp (1956)
and later emphasized by Capaccioli et al. (1989) is in fact real,
it is perhaps surprising that the absolute magnitude distribution for
our much larger sample of M31 novae shown in Figure~5 shows
no evidence for bimodality.
It may be possible to attribute this discrepancy to the fact that
many of the M31 novae in our sample likely had
faded somewhat by the time of discovery, in which case,
as noted earlier,
the absolute magnitudes used to construct Figure~5
do not all correspond to maximum light. Nevertheless,
given the likelihood
that the majority of novae were observed near peak brightness,
it would seem reasonable to expect some trace of bimodality
in the distribution to remain.

Another approach to testing for the existence of
two distinct nova populations is to compare the absolute magnitude
distribution of the M31 novae
with similar distributions for the Galaxy and the LMC.
Novae in M31 are
thought to arise mainly from the galaxy's bulge population
(Ciardullo et al. 1998, Cappacioli et al. 1989, Shafter \& Irby 2001,
Darnley et al. 2006), while those in the Galaxy and in the LMC
are associated primarily with a younger stellar population.
Although Galactic novae occur both in the bulge and disk,
the observed nova sample is
biased by relatively nearby objects, which are mostly
members of the disk population. Similarly,
novae in the LMC have been found to
belong primarily to a young population, with
progenitors estimated to be in the age range of $1-3.2$~Gyrs
(Subramaniam \& Anupama 2002). If novae arising from a younger
stellar population are on average more luminous than their
bulge counterparts, this difference should be reflected
in the luminosity distributions for the three galaxies.

To test this possibility, absolute magnitude distributions for
Galactic and LMC novae have been included in Figure~5
for comparison with the M31 distribution.
The Galactic data come from directly the compilations
of Downes \& Duerbeck (2000) and Shafter (1997), while
the LMC nova luminosities have been computed from
the apparent magnitudes given in Shida \& Liller (2004)
assuming $\mu_0(\rm{LMC})=18.5$ and $A_V=0.4$ (van den Bergh 2000).
Although the Galactic and M31 magnitude
distributions agree reasonably well at the bright end as expected, the
mean absolute magnitudes of the Galactic and LMC samples
($<$$M_V$$>=-7.79\pm0.11$ and $<$$M_V$$>=-8.06\pm0.05$, respectively)
are significantly brighter than the mean
of the M31 sample ($<$$M_V$$>=-7.20\pm0.04$).
As just alluded to with regard to the lack of bimodality in the overall
magnitude distribution, it is likely that some
of this discrepancy can be attributed to the fact that many
M31 novae have faded somewhat from maximum light by the time of discovery.
In addition,
extinction internal to M31, for which we have made no correction,
will also lower our estimates of the M31 nova luminosities.
However, it is also possible that part of the discrepancy is real, and
that it results from the differences in the
underlying nova populations.
In this case, since
the Hubble type of M31 (Sb) is significantly earlier than that of
the LMC (Irr~I) and slightly earlier than that of the Galaxy (Sbc),
M31 would be expected to contain a greater fraction of bulge novae
compared with the latter galaxies. 

\subsection{Properties of Luminous Novae}

In Tables 4, 5, and~6 we summarize properties of novae
in the Galaxy, M31, and the LMC, that have reached absolute
visual magnitudes brighter than $M_V=-9.0$. In addition to
their absolute magnitudes, we have listed, when available,
data for the rise time to maximum light, $t_R$,
the time to fade by two magnitudes from maximum light, $t_2$,
and the spectroscopic class. Novae exhibit a well-known
relationship between their peak luminosity
and their rate of decline known as the maximum-magnitude
vs. rate-of-decline (MMRD) relation (McLaughlin 1945,
Della Valle \& Livio 1995, Downes \& Duerbeck 2000). 
In his discussion of Nova LMC~1991,
Della Valle (1991) pointed out that several novae,
in addition to Nova LMC~1991 reached peak luminosities
about a magnitude brighter than than expected from
the mean MMRD relation. He concluded that a distinct
``super-bright" population of novae may exist.

To explore
this possibility further, and to see where M31N~2007-11d
fits into this picture, in Figure~6 we have plotted the
peak luminosities ($M_V[max]$) and corresponding fade rates
($t_2$) for our samples of luminous Galactic, M31 and LMC novae.
Since our samples are restricted to $M_V<-9.0$, the plot
only shows the bright end of the MMRD relation (i.e., less
luminous novae lying in the shaded region are not shown).
To better assess the fit to the full MMRD relation,
the reader is referred to the studies by Della Valle \& Livio (1995)
and Downes \& Duerbeck (2000).
Despite its limitations, Figure~6 reveals that
there is considerable scatter at the bright
end of the luminosity function with little correlation of
the maximum magnitude with $t_2$. Both M31N~2007-11d and
Nova LMC~1991 are close to a magnitude brighter than
predicted by the non-linear MMRD relation of Della Valle \& Livio (1995)
(less so for the linear relation of Downes \& Duerbeck [2000]
for Galactic novae), but it is not clear that they form part of a
separate class of super-bright novae. This view is supported
by the luminosity distribution shown in Figure~5, where no evidence
for a separate class is seen. Instead, we argue that
the scatter seen in Figure~6 results from a
combination of observational uncertainties (e.g., in extinction
corrections) coupled with the intrinsic variability one
expects from a population of novae with a distribution of
fundamental properties (e.g., white dwarf masses, mass accretion rates,
metallicites, etc.).

As mentioned earlier, there is evidence that the LMC
novae fade more quickly (i.e. have a lower mean
$t_2$) compared with Galactic and M31 novae.
The data for luminous novae from Tables~4--6 appear to be
consistent with this picture with
$<$$t_2$$>=6.0\pm4.4$ days, $<$$t_2$$>=10.6\pm3.8$ days,
and $<$$t_2$$>=3.7\pm1.7$ days,
for the Galaxy, M31, and the LMC respectively.
Furthermore, referring to Table~6, we see that
in the LMC a total of four novae (perhaps five if V2434-LMC
is included) have $M_V\lessim-9.0$. This number is quite high
given that a total of only $\sim34$ novae have been
discovered in the LMC (Shida \& Liller 2004). Thus, luminous
novae ($M_V\lessim-9.0$) appear to
make up roughly 12\% of the total.
In contrast, only a dozen ($\sim9$\%) out of the 127 Galactic novae
with measured luminosities ($\sim$5\% of all Galactic novae),
and only 26 out of the 732 M31 novae
($\sim4$\%) have similarly high luminosities. 

Finally, it is interesting to note that although Fe~II novae make up the
vast majority ($\sim$80\%) of all Galactic novae (Shafter 2007b), only three
of the 12 Galactic novae with $M_V\leq-9$ are members of the Fe~II class.
The remaining four systems are either He/N novae, or hybrid (Hy) objects
thought to be related to the He/N systems.
The LMC data, while limited, support the conclusion that
the He/N novae are overrepresented among the luminous novae.
Spectroscopic data for luminous novae in M31 are also limited,
but preliminary data again suggest that He/N novae tend to
be slightly more luminous that the Fe~II systems (Shafter et al. 2008,
in preparation). In M31 generally, Shafter (2007a) found that
$\sim15$\% of M31 novae with measured spectra
fall into the He/N class, which is comparable to, but
slightly smaller than the percentage seen in the Galaxy.
If this difference is found to be
statistically significant, it is possible that
the slightly higher fraction of He/N
novae observed in the Galaxy (and possibly in the LMC)
is reflecting either the selection bias
in favor of ``disk" novae, or the slightly later
Hubble types of the Milky Way and the LMC compared with M31, or both.

\subsection{The Rise Times of Luminous Novae}

In addition to its classification as an unusually luminous
Fe~II nova, M31N~2007-11d may be atypical in another respect.
In comparison with most Galactic novae,
the timescale of its rise to maximum light ($\grtsim4$~days)
appears at first glance
to be atypically slow for such a luminous nova.
Although observations of the rise to maximum in Galactic novae
are fragmentary, the existing observations
seem to indicate that essentially all novae reach maximum
light within three days (Payne-Gaposchkin 1957, Warner 1995, 2008).
For the fastest (and most luminous) novae, the rise to maximum
is believed to be less than two days, and in the case of
V1500 Cyg, may have been less than a day (Liller et al. 1975).
The scarcity of pre-maximum data can be attributed
to the rapid rise, which results in most novae
being discovered either at, or more typically shortly after, maximum light.
Strictly speaking, the rise time estimates given in Table~4
are best considered lower limits
since pre-maximum observations do not often extend more than a few magnitudes
below maximum light.
Although the statistics are poor, it is interesting to note that
of the seven novae for which rise times can be constrained,
the two Fe~II novae appear to have taken the longest time to
reach peak brightness.
It is tempting to speculate that the similarly
slow rise to maximum
observed in M31N~2007-11d might be related to its Fe~II nature.
 
The dearth of reliable data on the rise times of Galactic
novae is in part a result of how these objects are discovered:
usually serendipitously, often in patrols
for solar system bodies such as comets and asteroids.
On the other hand, most M31 and LMC novae have been discovered in
targeted surveys, some of which may have sufficiently frequent
temporal sampling to better constrain a nova's rise time to peak brightness.
Referring to Tables~5 and~6 we see that
sufficient data are available for about half of these novae
to establish that rise times of up to four days are perhaps not
as uncommon as the Galactic data would suggest.

One nova in particular, Nova LMC~1991, stands out as being remarkably
similar to M31N~2007-11d. Nova LMC~1991 is the brightest nova
to be observed in the LMC (and one of the brightest extragalactic
novae on record), reaching $M_V\simeq-10.1$. Despite its high luminosity,
Nova LMC~1991 took an usually long time to reach maximum light
($t_R\simeq10$~days); and, like M31N~2007-11d, has been classified
as a member of the Fe~II spectroscopic class (Williams et al. 1994).
Schwarz et al. (2001) have modeled the spectral evolution of
of Nova LMC~1991 during eruption and concluded that the luminosity
was super-Eddington before visual maximum, and that the ejected
mass was $\sim3\times10^{-4}$~M$_{\odot}$, which is about
an order of magnitude larger than that normally observed in ``fast"
novae. They argued that the high luminosity and large
ejected mass resulted from
accretion of low metallicity material onto a relatively cool,
high-mass white dwarf. The low metallicity allows more
material to be accumulated before a TNR is triggered than would
otherwise be possible for accretion onto a massive white dwarf.
The result is an energetic eruption characterized by a high shell
mass and a relatively slow photometric evolution.
It is plausible that M31N~2007-11d represents a similar system.

In view of available evidence, we conclude
that the rise time of $\sim$4 days observed
in M31N~2007-11d may not be as atypical as initially assumed, and
is likely the result of a relatively high ejected mass.
Unfortunately, with the exceptions of M31N~2007-11d and
Nova LMC~1991, few spectroscopic
classifications are available for other luminous novae
with well-determined rise times. Thus, at present,
we cannot draw any firm conclusions regarding possible correlations
between spectroscopic class and the rise time to maximum light.

\section{Conclusions}

Our principal conclusions can be summarized as follows:

1) M31N~2007-11d was one of the more luminous novae to be
discovered in M31 over the past century. After a relatively
slow rise to maximum light, the nova
reached $R=14.9$ on 20~Nov~2007~UT.

2) Extensive photometry obtained on the subsequent decline from
maximum light showed that the nova could be classified
as a relatively ``fast" nova, that took $\sim9.5$ and $\sim13$ days
to by decline 2 and 3 magnitudes from maximum light, respectively.

3) Spectroscopic observations obtained the day following
maximum light, and then again $\sim2$ and $\sim3$ weeks
post maximum reveal relatively narrow (FWHM $\sim2300$~km~s$^{-1}$)
Balmer and Fe~II emission
lines that clearly establish M31N~2007-11d as a member
of the Fe~II class in William's (1992) system.

4) Among the most luminous Galactic novae with $M_V\leq-9$,
only one in four is of the Fe~II type despite the fact that
$\sim80$\% of all Galactic novae are members of this class.
Thus, M31N~2007-11d appears to be relatively rare in this respect.

5) The photometric and spectroscopic properties of M31N~2007-11d
appear to be remarkably similar to that of another extragalactic
nova, Nova LMC~1991 (Della Valle 1991, Schwarz et al. 2001).
In particular, both were Fe~II novae that reached absolute magnitudes
brighter than $M_V=-9$, and were characterized by
relatively slow rises to maximum light.

6) Although there is considerable scatter at the upper end of the
MMRD relation, we find no compelling evidence for a {\it distinct} class of
super-bright novae. The large scatter in luminosity
among novae with $M_V\lessim-9.0$ in Figure~6, coupled
with the smooth drop-off in the luminosity distribution
in Figure~5, is consistent with the uniform variation expected
from a combination of observational uncertainties and
intrinsic variability in fundamental nova parameters.

7) The rise time to maximum light for Galactic novae is
poorly constrained, but has been generally regarded to be
of order one to two days for the most luminous novae. A
comparison with available data for luminous M31 and LMC novae
such as M31N~2007-11d and Nova LMC~1991 shows
rise times up to 4 days may not be unusual, and suggests that
Galactic nova rise times may have been underestimated.


\acknowledgments

We thank the referee, M. Della Valle, for his comments
on the original manuscript, and for bringing the properties
of Nova LMC~1991 to our attention. The work presented here is
based in part on observations obtained with the Marcario
Low Resolution Spectrograph on the Hobby - Eberly Telescope,
which is operated by McDonald Observatory on behalf of the University
of Texas at Austin, the Pennsylvania State University,
Stanford University, the Ludwig-Maximillians-Universitaet,
Munich, and the George-August-Universitaet, Goettingen.
Public Access time is available on the Hobby - Eberly Telescope
through an agreement with the National Science Foundation.
The Liverpool Telescope is operated on the island of
La Palma by Liverpool John Moores University in the Spanish Observatorio del
Roque de los Muchachos of the Instituto de Astrofisica de Canarias with
financial support from the UK Science and Technology Facilities Council
(STFC). FTN is operated by the Las Cumbres Observatory Global Telescope
network. Data from FTN was obtained as part of a joint programme
between Las Cumbres Observatory and Liverpool JMU Astrophysics Research
Institute. Support from NSF grants AST-0607682 (AWS) and AST-0707769 (RMQ)
is gratefully acknowledged. 

{\it Facilities:} \facility{HET}, \facility{Keck:I}, \facility{Liverpool:2m}, \facility{PO:1.5m}.

\clearpage

\begin{deluxetable}{lcccc}
\tablewidth{0pt}
\tablecolumns{5}
\tablecaption{Summary of Palomar 60-in Photometric Observations}
\tablehead{\colhead{$\mathrm{MJD}\tablenotemark{a}-54400$} & \colhead{$g$} & \colhead{$r$} & \colhead{$i'$} & \colhead{$z'$}}
\startdata
$24.416\pm0.026$ & $15.29\pm0.02$ & $15.20\pm0.02$ & $15.22\pm0.02$ & $15.34\pm0.03$ \\
$25.082\pm0.008$ & $15.46\pm0.02$ & $15.23\pm0.02$ & $15.23\pm0.02$ & $15.31\pm0.03$ \\
$25.164\pm0.003$ & $   \dots    $ & $15.25\pm0.02$ & $15.26\pm0.02$ & $15.35\pm0.03$ \\
$25.237\pm0.003$ & $15.55\pm0.02$ & $15.25\pm0.02$ & $15.23\pm0.02$ & $15.32\pm0.03$ \\
26.094         &    \dots       &   \dots        &   \dots        & $15.50\pm0.03$ \\
26.178         &    \dots       &   \dots        &   \dots        & $15.46\pm0.03$ \\
$26.352\pm0.001$ & $16.03\pm0.07$ & $15.66\pm0.06$ & $15.56\pm0.02$ & $15.49\pm0.03$ \\
$27.439\pm0.002$ & \dots          & $15.91\pm0.03$ & $16.05\pm0.03$ & \dots          \\
$33.100\pm0.003$ & $16.75\pm0.03$ & $16.45\pm0.04$ & $16.69\pm0.04$ & $16.08\pm0.04$ \\
$34.111\pm0.003$ & $16.92\pm0.03$ & $16.51\pm0.05$ & $16.78\pm0.04$ & $16.26\pm0.04$ \\
$37.073\pm0.003$ & $18.03\pm0.08$ & $17.12\pm0.06$ &  \dots         & $16.77\pm0.04$ \\
$39.071\pm0.003$ & $20.30\pm0.10$ & $19.13\pm0.08$ & $19.12\pm0.07$ & $18.27\pm0.06$ \\
$40.211\pm0.003$ & $21.24\pm0.12$ & $20.10\pm0.10$ & $20.28\pm0.10$ & $19.05\pm0.07$ \\
$45.090\pm0.004$ & $>21.5$ & $>21.5$ & $>21.3$ & $20.96\pm0.12$ \\
46.096         & \dots   & \dots   & \dots   & $>21.2$ \\
\enddata
\tablenotetext{a}{$\mathrm{MJD} \equiv \mathrm{HJD}-2,400,000.5$.
The date given is the mean time for measurements in the different filters,
with the uncertainty representing the range in times.}
\end{deluxetable}

\begin{deluxetable}{lcccc}
\tablewidth{0pt}
\tablecolumns{5}
\tablecaption{Summary of FTN and LT Photometric Observations}
\tablehead{\colhead{$\mathrm{MJD}-54400$} & \colhead{Telescope} & \colhead{$V$} & \colhead{$B$} & \colhead{$i'$}}
\startdata
27.38 & FTN & $16.46\pm0.03$ & $16.07\pm0.02$ & $15.78\pm0.01$ \\
28.28 & FTN & $16.52\pm0.04$ & $16.24\pm0.03$ & $15.89\pm0.03$ \\
29.28 & FTN & $16.56\pm0.03$ & $16.33\pm0.02$ & $16.04\pm0.01$ \\
31.97 & LT  & $16.64\pm0.03$ & $16.60\pm0.02$ & $16.31\pm0.01$ \\
36.09 & LT  & $17.56\pm0.03$ & $17.56\pm0.03$ & $17.07\pm0.01$ \\
36.28 & FTN & $17.82\pm0.07$ & $17.70\pm0.04$ & $17.27\pm0.03$ \\
38.87 & LT  & $20.22\pm0.04$ & $20.04\pm0.04$ & $18.71\pm0.02$ \\
42.84 & LT  & $21.77\pm0.03$ & $21.44\pm0.13$ & $20.27\pm0.07$ \\
48.84 & LT  & \dots          & \dots          & $20.88\pm0.09$ \\
\enddata
\end{deluxetable}

\begin{deluxetable}{lcc}
\tablewidth{0pt}
\tablecolumns{3}
\tablecaption{ROTSE IIIb Photometry}
\tablehead{\colhead{$\mathrm{MJD}-54400$} & \colhead{Unfiltered Mag} & \colhead{Limiting Mag}}
\startdata
15.08 &  \dots          & 18.79\\
16.06 &  \dots          & 18.66\\
17.13 &  \dots          & 19.17\\
18.06 &  \dots          & 19.09\\
20.06 &  $17.80\pm0.25$ & 17.78\\
21.06 &  $17.24\pm0.11$ & 18.91\\
23.08 &  $15.76\pm0.03$ & 18.95\\
24.06 &  $15.30\pm0.02$ & 18.45\\
25.06 &  $15.42\pm0.03$ & 17.96\\
27.08 &  $16.03\pm0.05$ & 18.00\\
28.06 &  $16.18\pm0.06$ & 18.03\\
31.07 &  $16.75\pm0.22$ & 16.53\\
32.07 &  $16.46\pm0.06$ & 18.86\\
33.10 &  \dots          & 11.87\\
34.06 &  $17.06\pm0.12$ & 17.68\\
36.06 &  $17.39\pm0.19$ & 17.39\\
37.06 &  $17.52\pm0.15$ & 19.16\\
38.06 &  $17.70\pm0.43$ & 18.13\\
39.06 &  $18.47\pm0.53$ & 18.89\\
40.06 &  \dots          & 19.05\\
42.07 &  \dots          & 19.02\\
43.06 &  \dots          & 18.67\\
\enddata
\end{deluxetable}

\begin{deluxetable}{lccccc}
\tablewidth{0pt}
\tablecolumns{6}
\tablecaption{Properties of Luminous Galactic Novae with $M_V\lessim-9.0$}
\tablehead{\colhead{Nova} & \colhead{Spectral Class} & \colhead{$t_R$ (d)} & \colhead{$t_2$ (d)} & \colhead{$M_V$} & \colhead{References\tablenotemark{a}}}
\startdata
V500 Aql  & He/N? &\dots&  17  & $-9.0$  & 1 \\
V603 Aql  & Hy    &$\grtsim1$ &   4  & $-9.0$  & 1--4 \\
V476 Cyg  & Fe~II &$\grtsim4$ &   6  & $-9.9$  & 1,2,5 \\
V1500 Cyg & Hy    &$\sim1$ &  2.4 & $-10.7$ & 1,2,6 \\
V446 Her  & He/N  &\dots&   7  & $-9.9$  & 1,2 \\
CP Lac    & Hy    &$\grtsim2$ &  5.3 & $-9.3$  & 1,2,7 \\
DK Lac    & Fe~II &$\grtsim3$ &  11  & $-9.8$  & 1,2,8 \\
GK Per    & He/N  &$\grtsim1$ &   7  & $-9.0$  & 1,2,9 \\
CP Pup    & He/N  &$\grtsim2$ &   6  & $-10.7$ & 1,2,10 \\
V838 Her  & He/N  &\dots&  1.2 & $-9.0$  & 2,3,11 \\
V977 Sco  & Fe~II &\dots&  3.4\tablenotemark{b}   & $-9.0$ & 3,12 \\
MU Ser    & He/N  &\dots&  2  & $-9.0$ & 3,13  \\
\enddata
\tablenotetext{a}{(1) Downes \& Duerbeck (2000);
(2) Della Valle \& Livio (1998); (3) Shafter (1997);
(4) Duerbeck (1987); (5) Denning (1920); (6) Liller (1975);
(7) Bertaud (1945); (8) Bertaud (1950); (9) Campbell (1903);
(10) Dawson (1942); (11) Leibowitz (1993); (12) Liller (1993);
(13) Schlegel et al. (1985)}
\tablenotetext{b}{Estimated from $t_3$ using the relationship,
$t_2\simeq(t_3/2.75)^{1.14}$ from Warner (1995).}
\end{deluxetable}

\begin{deluxetable}{lccccc}
\tablewidth{0pt}
\tablecolumns{6}
\tablecaption{Properties of Luminous M31 Novae with $M_V\lessim-9.0$}
\tablehead{\colhead{Nova} & \colhead{Spectral Class} & \colhead{$t_R$ (d)} & \colhead{$t_2$ (d)} & \colhead{$M_V$} & \colhead{References\tablenotemark{a}}}
\startdata
M31N 1925-09a & \dots & $\grtsim1 $ & $ >8 $   & $-9.3$  & 1 \\
M31N 1932-07a & Fe~II & $\grtsim4 $ & $ 10 $   & $-9.1$  & 2 \\
M31N 1960-11a & \dots & $\grtsim2 $ & $ 7 $   & $-9.7$  & 3 \\
M31N 1963-09b\tablenotemark{b} & \dots & \dots & \dots & $-9.5$  & 3 \\
M31N 1964-11b & \dots & $\grtsim3 $ & $ >7$  & $-9.6$  & 3 \\
M31N 1965-12a & \dots & $\grtsim4 $ & $ <6$  & $-9.3$  & 3 \\
M31N 1967-10c & \dots & \dots & $>11$   & $-9.6$ & 4 \\
M31N 1975-02a & \dots & $\grtsim1 $ & \dots   & $-9.2$ & 4 \\
M31N 1981-09b & \dots &\dots& \dots  & $-9.3$  & 5 \\
M31N 1981-09c & \dots &\dots& \dots  & $-9.2$  & 5 \\
M31N 1982-08b & \dots &$\grtsim1$ & $<13$ & $-9.2$  & 6 \\
M31N 1983-01a\tablenotemark{c} & \dots &\dots& $<3.3$ & $-9.5$  & 6--8 \\
M31N 1983-09d\tablenotemark{d} & \dots &$>32$&\dots  & $-10.5$  & 9 \\
M31N 1983-10a\tablenotemark{d} & \dots &$>3$ &\dots  & $-10.1$  & 9 \\
M31N 1983-10b\tablenotemark{d} & \dots &\dots&\dots  & $-9.4$  & 9 \\
M31N 1984-10b\tablenotemark{d} & \dots &\dots&\dots  & $-9.0$  & 9 \\
M31N 1985-10c\tablenotemark{d} & \dots &$>3$ &\dots  & $-9.2$  & 9 \\
M31N 1990-12c & \dots & \dots & $>2$    & $-9.7$  & 6 \\
M31N 1991-01a & \dots & \dots & \dots   & $-9.7$  & 10 \\
M31N 1993-01a\tablenotemark{d} & \dots &$>3$ &\dots  & $-9.1$  & 11 \\
M31N 1995-12a & \dots & \dots       & \dots  & $-9.4$ & 12 \\
M31N 1996-08g & \dots & $\grtsim3 $ & $>3$   & $-9.7$ & 13 \\
M31N 1998-07n\tablenotemark{e} & \dots & $\grtsim5 $ & $>4$  & $-10.0$  & 14,15 \\
M31N 2004-09b & \dots & \dots       & \dots   & $-9.3$ & 16 \\
M31N 2005-01a & \dots & $\grtsim3 $ & $16$   & $-9.4$ & 17 \\
M31N 2007-11d & Fe~II & $\grtsim4 $ & $9.5$   & $-9.5$ & 18 \\
\enddata
\tablenotetext{a}{(1) Hubble (1929); (2) Humason (1932); (3) Rosino (1973); (4) Henze at al. (2008); (5) Ciardullo et al. (1983); (6) Sharov \& Alksnis (1992b); (7) Bryan \& Brewster (1983); (8) Sharov \& Alksnis (1992a); (9) Ciardullo et al. (1987); (10) Birkle et al. (1991); (11) Shafter \& Irby (2001); (12) Ansari et al. (2004); (13) Sharov \& Alksnis (1997); (14) Modjaz et al. (1998); (15) Sharov et al. (2000); (16) Tzenev et al. (2004); (17) Hornoch et al. (2005); (18) This work.}
\tablenotetext{b}{Maximum magnitude is uncertain; based on extrapolation}
\tablenotetext{c}{Object questionable; not confirmed by Sharov \& Alksnis (1992)}
\tablenotetext{d}{Observations in H$\alpha$; the rise times likely do not
reflect the behavior in the continuum.}
\tablenotetext{e}{Magnitude at maximum light from Modjaz et al. (1998)
and Sharov et al. (2000) differ significantly, making $M_V$ particularly
uncertain.}
\end{deluxetable}

\begin{deluxetable}{lccccc}
\tablewidth{0pt}
\tablecolumns{6}
\tablecaption{Properties of Luminous LMC Novae with $M_V\lessim-9.0$}
\tablehead{\colhead{Nova} & \colhead{Spectral Class} & \colhead{$t_R$ (d)} & \colhead{$t_2$ (d)\tablenotemark{a}} & \colhead{$M_V$} & \colhead{References\tablenotemark{b}}}
\startdata
LMC 1978a\tablenotemark{c} & He/N? & \dots       & $3.5$  & $-9.2$ & 1,2 \\
LMC 1987  & \dots & \dots       & $2.0$  & $-9.3$ & 1,3 \\
LMC 1990a & He/N  & \dots  & $3.4$   & $-9.2$ & 1,4 \\
LMC 1991  & Fe~II & $\grtsim10 $ & $6.0$   & $-10.1$ & 1,5--10 \\
V2434-LMC\tablenotemark{d}  & \dots & \dots & $\lessim1$  & $-9.2$ & 1,11 \\
\enddata
\tablenotetext{a}{$t_2$ values from Della Valle (1991)}
\tablenotetext{b}{(1) Shida \& Liller (1991); (2) Graham (1979); (3) McNaught \& Garradd (1987); (4) Dopita \& Rawlings (1990); (5) Della Valle (1991); (6) Gilmore \& Liller (1991); (7) Della Valle et al. (1991); (8) Dopita et al. (1991); (9) Wiliams et al. 1994; (10) Schwarz et al. (2001); (11) Liller \& Morel (2002)}
\tablenotetext{c}{Maximum magnitude is uncertain; based on extrapolation}
\tablenotetext{d}{Nature of object uncertain; possible nova}
\end{deluxetable}

\clearpage




\begin{figure}
\includegraphics[angle=0,scale=.65]{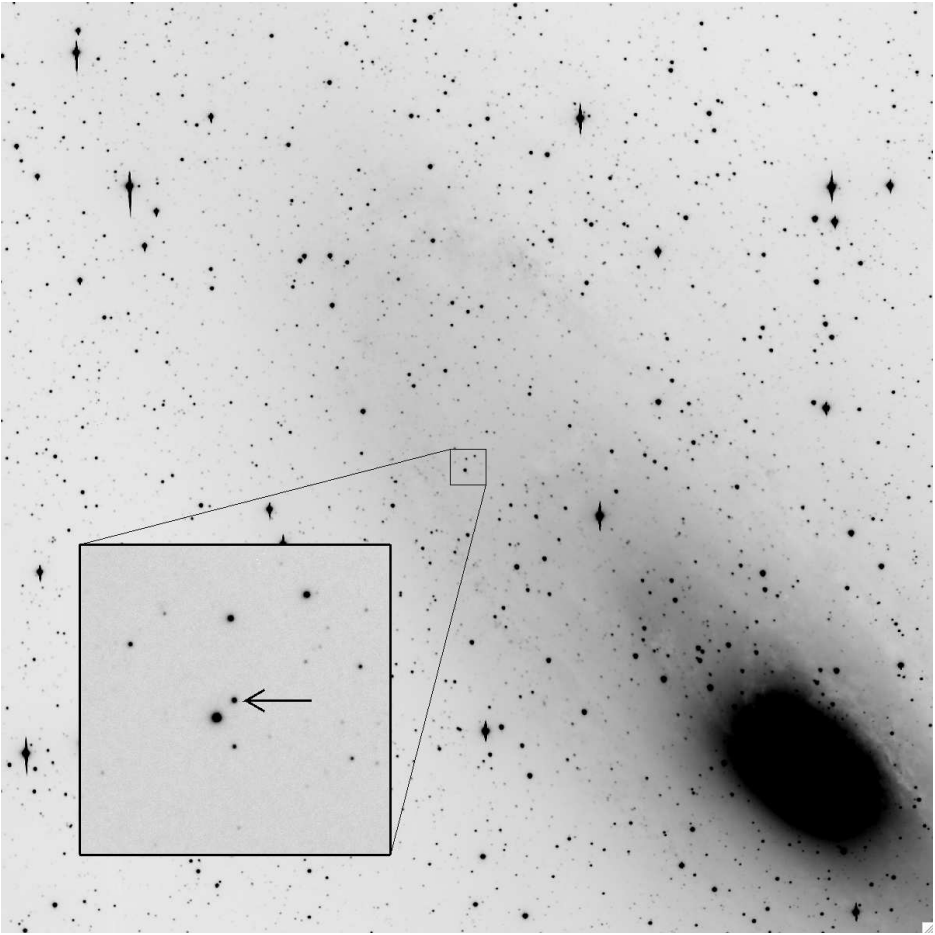}
\caption{Messier 31 as observed by ROTSE-IIIb and P60 showing the
location of M31N 2007-11d. The large image
is a co-addition of archival ROTSE-IIIb images, with
the inset ($\sim2.5'\times2.5'$ on a side) based on a P60 image
from 2007 November 21. North is up and East is to the left.
\label{fig1}}
\end{figure}

\begin{figure}
\includegraphics[angle=-90,scale=.65]{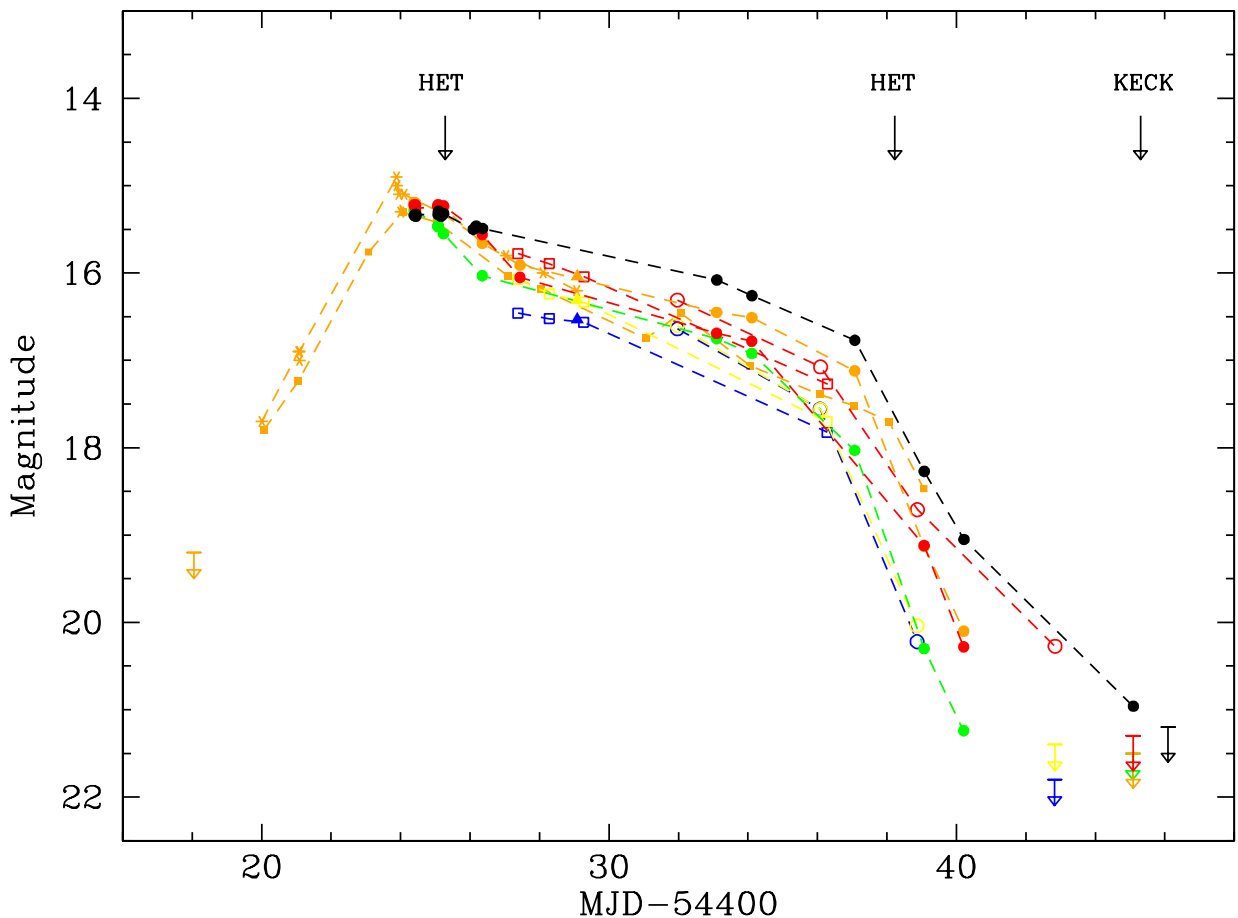}
\caption{The light curve of M31N2007-11d. Key:
P60 -- filled circles ($g$ -- green, $r$ -- orange, $i'$ -- red,
and $z'$ -- black), FTN -- open squares ($B$ -- blue, $V$ -- yellow,
$i'$ -- red), LT -- open circles ($B$ -- blue, $V$ -- yellow,
$i'$ -- red), MLO -- filled triangles ($B$ -- blue, $V$ -- yellow,
$R$ -- orange). Also shown are
pre-maximum measurements from K. Nishiyama and F. Kabashima
(orange asterisks) and ROTSE-IIIb (orange filled squares).
The times of our spectroscopic observations are indicated by the arrows.
Note the relatively slow rise to maximum, and the
rapid diminution in brightness near the time of our second HET spectrum.
\label{fig2}}
\end{figure}

\begin{figure}
\includegraphics[angle=-90,scale=.65]{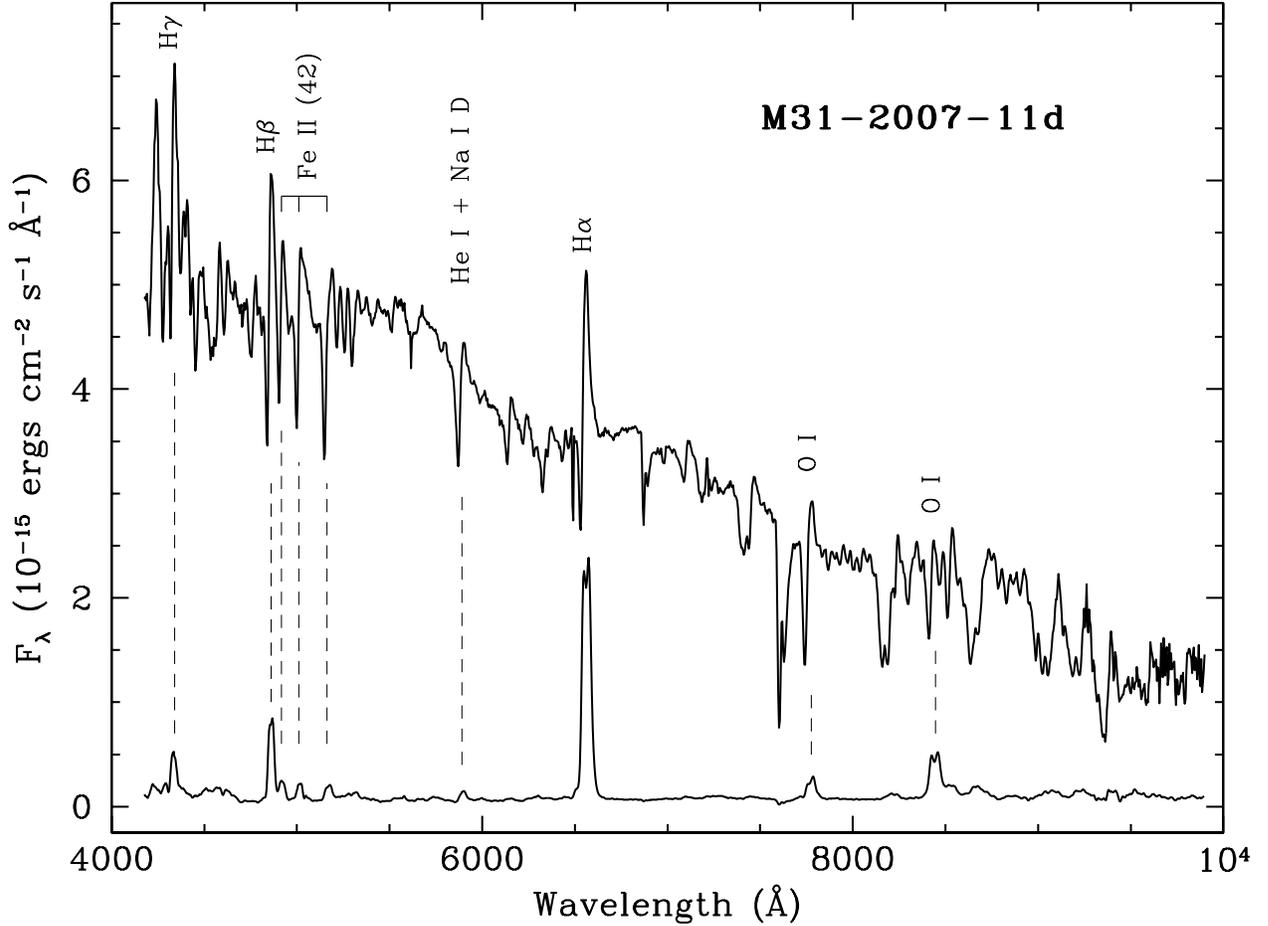}
\caption{Low-resolution spectra of M31N2007-11d taken with the LRS on the
HET $\sim5$ days after discovery and near maximum light (upper spectrum),
and $\sim18$~days after discovery (lower spectrum).
The initial spectrum is characterized by Balmer and Fe~II emission
with P~Cygni profiles.
As the nova faded, the spectrum evolved into a standard
Fe~II type.\label{fig3}}
\end{figure}

\begin{figure}
\includegraphics[angle=-90,scale=.65]{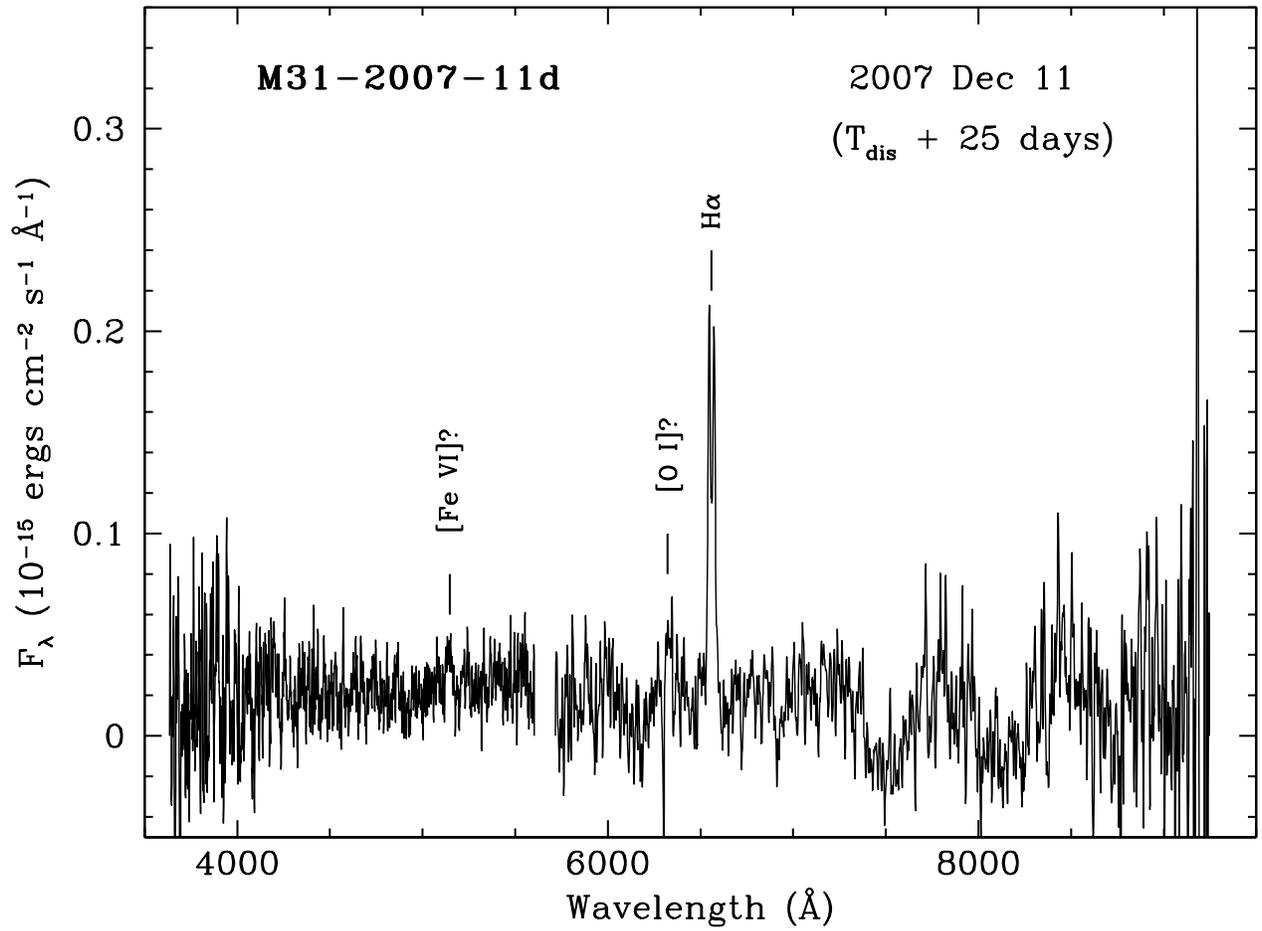}
\caption{A low-resolution spectrum taken with the Keck/LRIS
$\sim25$ days post-discovery when the nova had faded significantly.
\label{fig4}}
\end{figure}

\begin{figure}
\includegraphics[angle=-90,scale=.65]{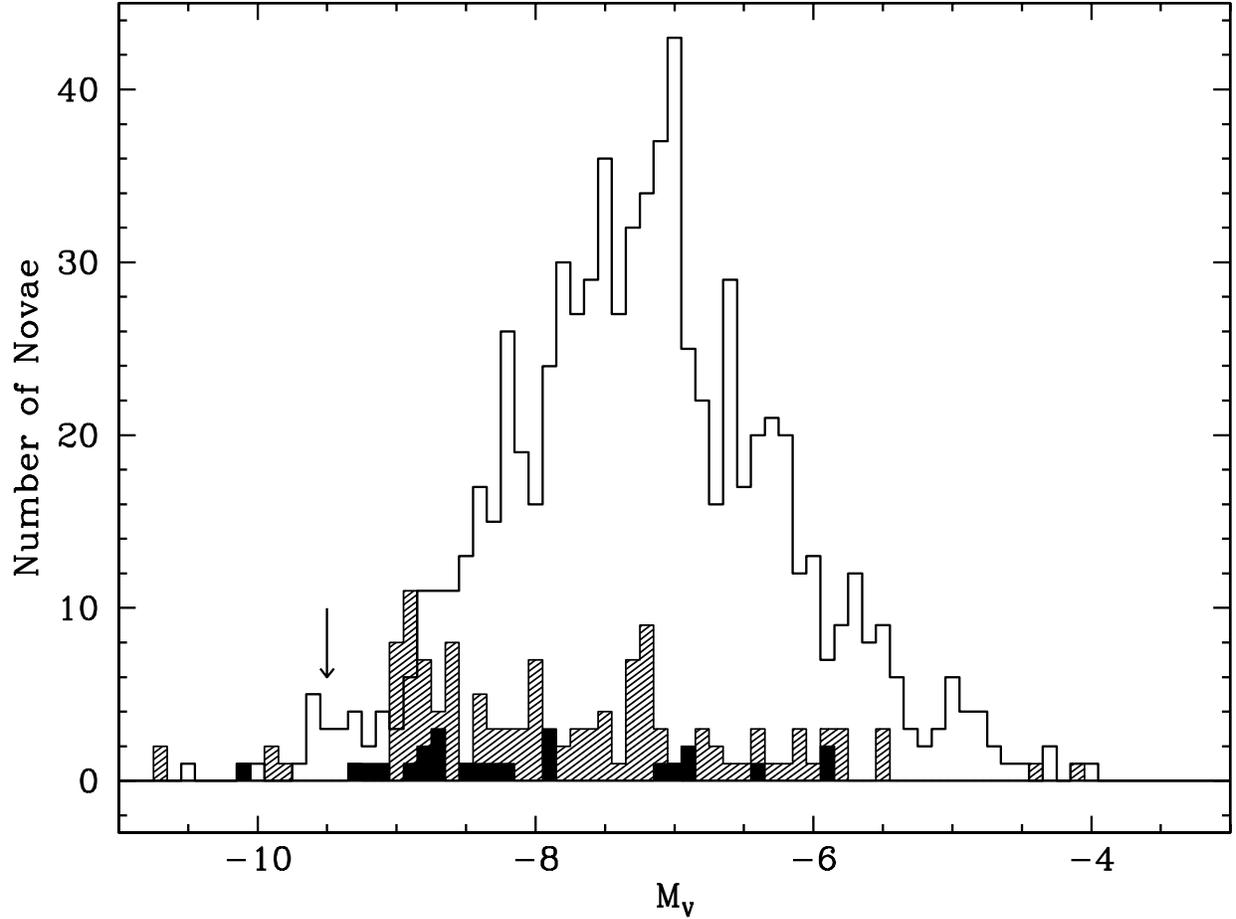}
\caption{The absolute visual magnitude distribution of the 732 M31
novae after making corrections for foreground
extinction, bandpass of observation,
and distance to M31. For comparison,
the cross-hatched and filled regions show the
$M_V$ distributions for 127 Galactic novae and 24 LMC novae,
respectively.
The arrow shows the position of M31N~2007-11d.
\label{fig5}}
\end{figure}

\begin{figure}
\includegraphics[angle=-90,scale=.65]{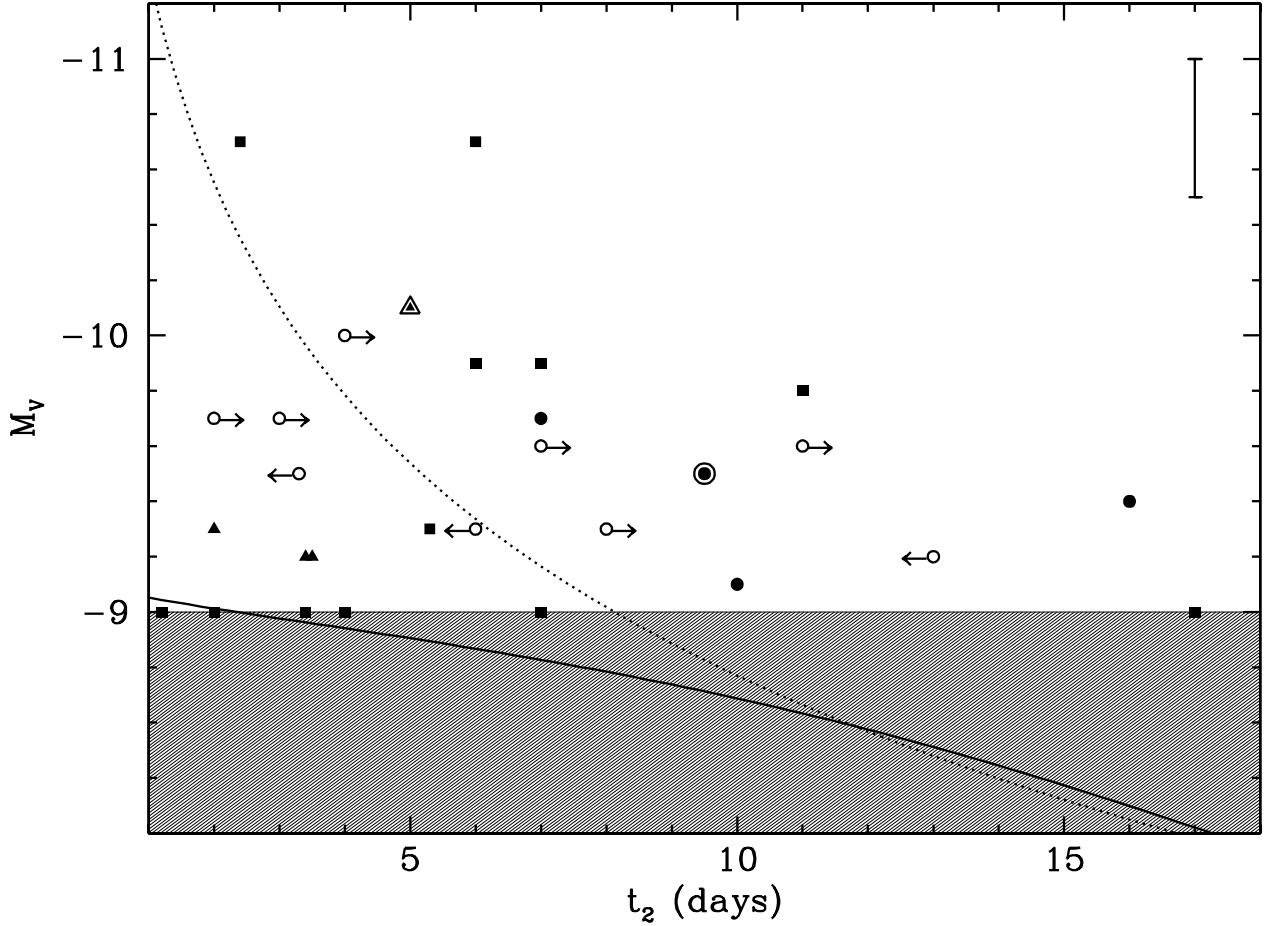}
\caption{The absolute magnitudes of novae
with $M_V\lessim-9.0$ taken from Tables~4, 5, and~6
are plotted as a function of the time to fade by two magnitudes, $t_2$.
Novae lying in the shaded region (below our luminosity cutoff) are not plotted.
The filled squares circles and triangles represent Galactic,
M31, and LMC novae, respectively. Open circles represent M31 novae with only
upper or lower limits available for $t_2$. The filled circle surrounded
by an open circle represents M31N 2007-11d, and for comparison
the similar object, Nova LMC~1991, is shown as a filled triangle
surrounded by an open triangle. The error bar in the upper right
shows the estimated uncertainty in $M_V$ for a typical nova.
The solid and dotted lines represent the MMRD relations of
Della Valle \& Livio (1995) and Downes \& Duerbeck (2000), respectively.
\label{fig6}}
\end{figure}

\end{document}